**POLICY MEMO
JANUARY 2025**

# The Future of the AI Summit Series

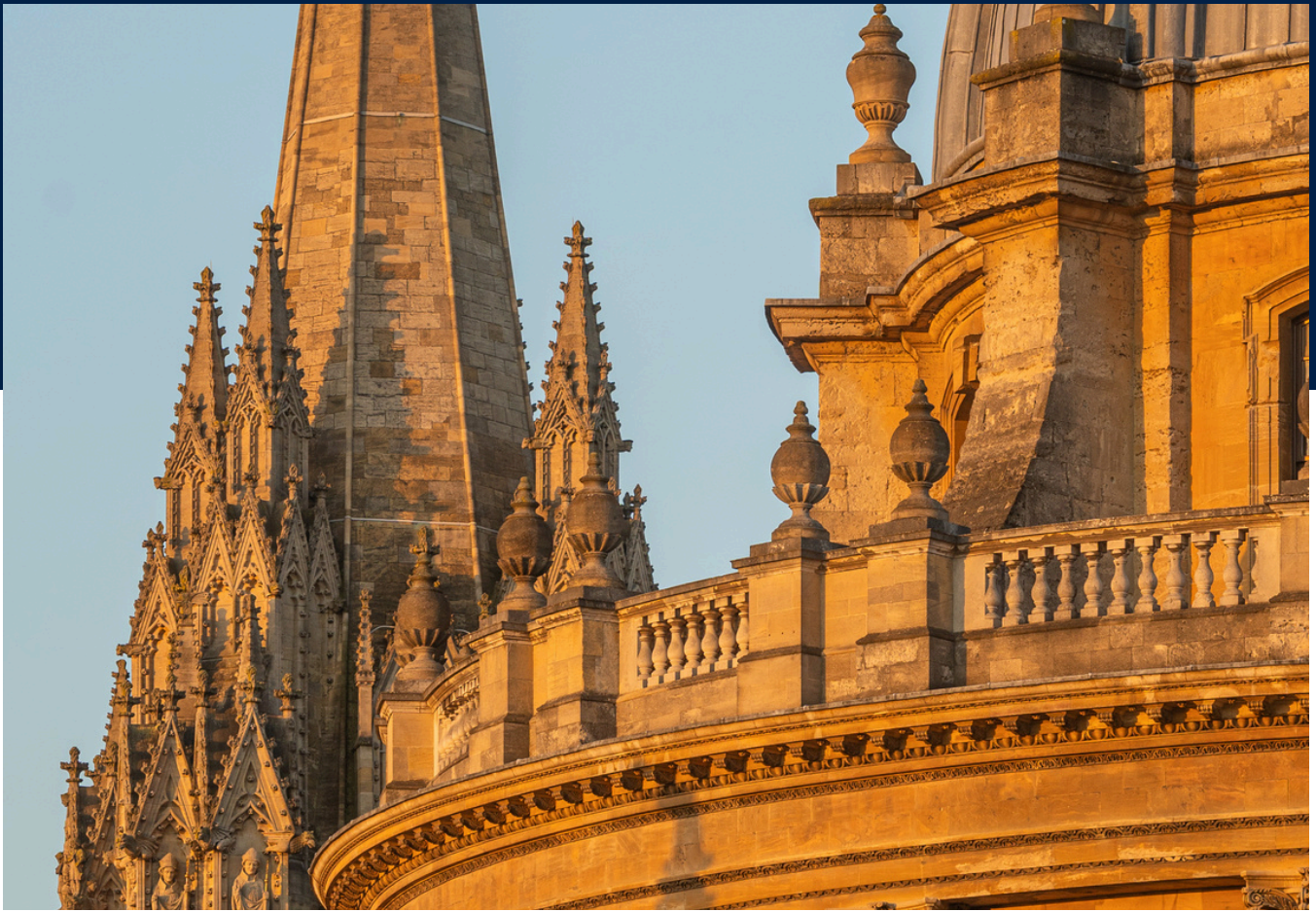

Authors: Lucia Velasco, Charles Martinet, Henry de Zoete, Robert Trager, Duncan Snidal, Ben Garfinkel, Kwan Yee Ng, Haydn Belfield, Don Wallace, Yoshua Bengio, Benjamin Prud'homme, Brian Tse, Roxana Radu, Ranjit Lall, Ben Harack, Julia Morse, Nicolas Miailhe, Scott Singer, Matt Sheehan, Max Stauffer, Yi Zeng, Joslyn Barnhart, Imane Bello, Xue Lan, Oliver Guest, Duncan Cass-Beggs, Lu Chuanying, Sumaya Nur Adan, Markus Anderljung, and Claire Dennis.

# The Future of the AI Summit Series

Lucia Velasco[1†*], Charles Martinet[1,2†*], Henry de Zoete[1†*], Robert Trager[1†**], Duncan Snidal[3*], Ben Garfinkel[3*], Kwan Yee Ng[4*], Haydn Belfield[5*], Don Wallace[6*], Yoshua Bengio[7,8], Benjamin Prud'homme[8], Brian Tse[4], Roxana Radu[3], Ranjit Lall[3], Ben Harack[1,3], Julia Morse[1], Nicolas Miailhe[9], Scott Singer[10], Matt Sheehan[10], Max Stauffer[11], Yi Zeng[12], Joslyn Barnhart[6], Imane Bello[13], Xue Lan[14], Oliver Guest[15], Duncan Cass-Beggs[16], Lu Chuanying[17], Sumaya Nur Adan[1], Markus Anderljung[18], Claire Dennis[18]

*[1] Oxford Martin AI Governance Initiative, [2] French Center for AI Safety (CeSIA), [3] University of Oxford, [4] Concordia AI, [5] University of Cambridge, [6] Google Deepmind, [7] University of Montréal, [8] Mila - Quebec AI Institute, [9] PRISM Eval, [10] Carnegie Endowment for International Peace, [11] Simon Institute for Longterm Governance, [12] Beijing Institute of AI Safety and Governance, [13] Future of Life Institute, [14] Institute for AI International Governance, Tsinghua University, [15] Institute for AI Policy and Strategy, [16] Centre for International Governance Innovation, [17] Cyberspace International Governance Research Center, Shanghai Institute for International Studies, [18] Centre for the Governance of AI*

Corresponding author: lucia.velasco.j@gmail.com

† Primary authors who contributed most significantly to the direction and content of the paper
\* Main co-authors
\*\* Senior author

We are grateful to Sam Daws, Joe Jones, Sumaya Nur Adan, and José Villalobos for their valuable feedback on this project.

Please cite as: Velasco, Trager, et al. (2025). "The Future of the AI Summit Series", *Oxford Martin School AI Governance Initiative Policy Memo*

# The Future of the AI Summit Series

This document is a policy briefing intended to inform decision-makers in governments, international organizations, and industry in upcoming decisions about the summit series. It is not an academic research paper, but rather a practical guide to help stakeholders identify key strategic considerations, weigh potential options, and decide on next steps for shaping the summit series.

This briefing is informed by research conducted by primary authors from the Oxford Martin AI Governance Initiative (Lucia Velasco, Charles Martinet, Henry de Zoete, and Robert Trager), main co-authors Duncan Snidal, Ben Garfinkel, Kwan Yee Ng, Haydn Belfield, Don Wallace, and secondary co-authors Yoshua Bengio, Benjamin Prud'homme, Brian Tse, Roxana Radu, Ranjit Lall, Ben Harack, Julia Morse, Nicolas Miailhe, Scott Singer, Matt Sheehan, Max Stauffer, Yi Zeng, Joslyn Barnhart, Imane Bello, Xue Lan, Oliver Guest, Duncan Cass-Beggs, Lu Chuanying, Sumaya Nur Adan, Markus Anderljung, and Claire Dennis. It is further enriched by insights gathered during an expert workshop hosted by the Initiative, where representatives from the organizers of previous Summits participated and provided valuable context. We also wish to acknowledge the important contributions of Sam Daws, Joe Jones, and José Villalobos, whose expertise and suggestions significantly improved this document.

Given the number of authors, authorship does not imply agreement with every recommendation made in this paper.

# Executive summary

The AI Summit series – initiated at Bletchley Park in 2023 and continuing through Seoul in 2024 and Paris in 2025 – has become a distinct forum for international collaboration on AI governance. Its early achievements, including the Bletchley Declaration, the Frontier AI Safety Commitments, and the International Scientific Report on the Safety of Advanced AI, are a result of its unique format, regular schedule, and ability to secure concrete commitments from governments and industry.

To ensure its continuing impact, the Summit series must now transition from an improvised sequence of summits towards a more formalized structure. For this evolution to succeed, organizers must carefully examine past successes and realistically assess future challenges. This report examines both, with particular attention to a set of core summit design elements: hosting arrangement, secretariat format, participant selection, agenda setting, and summit frequency. Based on this analysis, we present six recommendations to strengthen the summit series' impact.

The paper draws on existing international governance models to offer recommendations for each design element, addressing challenges such as a crowded summit landscape, geopolitical shifts, and rapid technological change. **Table 1** summarizes the elements and options. We assess each option for its potential to contribute to the series' long-term effectiveness and describe tradeoffs.

### Table 1: Overview of Design Options for the AI Summit Series

| Element | Option A | Option B | Option C | Option D | Recommendation |
|---|---|---|---|---|---|
| 1. Hosting Arrangement | **Rotation of hosting** Hosting passes between core members on a fixed schedule | **Bidding for hosting** Members vote among candidates | **Regional groups hosting** Hosting rotates among regions | **Joint hosting** Two or more countries share hosting duties | **Bidding system with regional rotation** *Mix of options B and C* |
| 2. Secretariat Format | **Light coordination** Host handles with a small working group | **Hybrid/incremental** Core team retained, host has leeway | **Formal secretariat** Fully independent from host, (hosted by intl. org) | N/A | **Hybrid/incremental** *Option B* |
| 3. Participant Selection | **Host-driven selection** Host country decides invitations | **Curated membership** Predetermined participation criteria | **Blended model** Core group + host-invited guests | **Universal participation** Open to all countries | **Two-track approach** *Mix of options B and C* |
| 4. Agenda Setting | **Host-led** Host country sets primary agenda | **Multiple workstreams** Parallel tracks run with different stakeholder groups | **Coordinated succession** Multiple hosts coordinate over extended period | **Steering committee** Intl. multi-stakeholder group sets priorities | **Coordinated succession** *Option C* |
| 5. Summit Frequency | **Annual summits with ongoing engagement** | **Flexible biannual summits** | **Annual summits with interim meetings** | N/A | **Annual summits with interim meetings** *Option C* |

**Recommendations**

A central recommendation following from the analysis is for the summit series to retain its **core focus on the international governance of advanced AI**. This focus addresses a gap in the current ecosystem, where no other forum is dedicated specifically to the most capable AI systems. We define "advanced AI" as *general-purpose or specialized AI systems that match or exceed the capabilities of the most powerful systems today*. These systems could pose unique risks and opportunities that require international attention. Thus, the series should make these systems its strategic priority, while recognizing the broader AI ecosystem's interconnected nature. We argue that this core focus on the governance of advanced AI should be maintained alongside a broader set of potential host-driven initiatives.

We propose a **two-track participation** model due to the need for both focused expertise and broad inclusivity. Track 1 will concentrate on the specific governance challenges of advanced AI, drawing on the expertise and technological leadership of a core group of leaders in the field. Track 2 will explore the opportunities AI brings and how to leverage it in the public interest, ensuring a more diverse range of voices.

**Core design elements**

To support this vision, we provide the following recommendations.

- **Meeting frequency:** Maintain the annual summit format, supplemented by interim meetings. Annual summits provide a regular forum for high-level political engagement, decision-making, and announcements. Interim technical-ministerial meetings allow participants to have focused discussions on timely issues and scientific advancements. Interim meetings allow participants to be responsive to the rapid pace of AI development, to track progress on prior summit commitments, and to prepare input for the main annual summit.

- **Secretariat:** Adopt a hybrid model with a semi-permanent core team retained across multiple summits to ensure modest institutional memory and stable operations, while allowing hosts the flexibility to shape agendas and invite participants. This could involve partial rotation, such as a small permanent unit complemented by a host-appointed team, striking a balance between continuity and adaptability. Over time, as the summit's scope broadens or membership grows, this structure could scale up its administrative resources and official mandates incrementally, avoiding the creation of a full-blown bureaucracy.

- **Participation:** We propose a two-track framework. Track 1 will focus on the governance of advanced AI, with attention to aspects such as cross-border risks, regulatory convergence and other geostrategic topics. Track 2 will provide a space to explore how advanced AI can serve the public interest[1], while examining its broader implications for society. For Track 1, criteria for participation must be established and reviewed periodically. Criteria could include, but are not limited to, national AI capabilities, jurisdiction over entities developing state-of-the-art AI, concentration of expert talent, and/or established or developing regulatory frameworks. We further propose adding the leading AI labs to this track. The inclusion of non-state actors, specifically leading AI companies and labs, is crucial given their role in developing advanced AI. If only countries are allowed to be "members" the private sector could have an observer status. Track 2 would include a broader set of countries invited by the host and/or proposed by the steering committee in charge of running the governance of the summits and composed of Track 1 members.

---

[1] Public interest AI refers to AI systems and technologies developed and deployed with the primary goal of serving the collective well-being and long-term interests of society

- **Hosting:** Adopt a selection model which combines a bidding system, where countries bid to host the summit and vote between the bids, with regional rotation when feasible. Track 1 participants – the core group of leading AI countries – would select and be eligible to bid and host. The selection would be finalized two years in advance to ensure continuity and preparation time, working under a "troika system"[2]. This approach provides a structured, adaptable process that reduces uncertainty.

- **Agenda Setting:** The "troika" will also contribute to a coordinated succession model where three consecutive hosts collaborate. We propose a multi-year roadmap, with a Track 1 committee to decide on the agenda. This committee will regularly update the roadmap with expert input to guide the summit series' direction. The Track 1 agenda will be decided by the countries which are part of the track, focusing on advanced AI governance. The host will set the Track 2 agenda, addressing broader public interest issues. Hosts retain final authority over their summit's deliverables.

- **Regular engagement with the United Nations and AISI Network:** Establish regular engagement with the United Nations to enhance the legitimacy and impact of the series and with the AISI Network to build on technical knowledge. This could involve mechanisms such as aligning meeting schedules, sharing information, and ensuring complementarity between the series' focus on advanced AI and the UN's broader digital governance mandate stemming from the Global Digital Compact.

---

[2] A troika refers to a model where the previous, current, and next host or chair form a coordinating group to ensure smooth transitions and institutional memory.

# 01.  Introduction

The landscape of international AI governance is being shaped by an emerging series of summits. The series began with the Bletchley AI Safety Summit in 2023, which was followed by the Seoul AI Summit in 2024, and it will continue with the upcoming Paris AI Action Summit in 2025. The first two summits have been successful, facilitating agreements on managing risks associated with advanced AI[3] and securing commitments from major companies and countries. Early achievements include the first international [declaration on advanced AI safety](), [voluntary commitments]() to implement safety frameworks from leading AI companies (from North America, Europe, Asia, China, and the Middle East), the commissioning of the [International Scientific Report on the Safety of Advanced AI](), and the establishment of national AI Safety Institutes, including in the [UK]() and [US]().

As we approach the third summit in Paris, however, we need to maintain the summits' momentum and relevance. Parallel efforts with varying scopes and memberships are emerging, which could cause political attention on the summit series to wane and its focus to drift.

The purpose of this memo is to inform decision-makers about critical strategic considerations for the future summit series, evaluate potential options, and recommend concrete next steps. Drawing upon an analysis of past summits and their key challenges, we propose actionable improvements across five core design elements: hosting arrangements, secretariat format, participant selection, agenda setting, and meeting frequency.

## 1.1. Drivers of summit series successes

The summit series has been successful in three key ways: it has carved out a unique niche that complements other initiatives in the crowded AI governance space; it has maintained a regular cadence without a formal structure; and it has driven action from governments and industry.

**Complementing other initiatives:** The summits address gaps left by broader initiatives like those of the OECD, GPAI[4], G7, or UN.

**Sustaining momentum:** The ability to convene regularly without a formal framework has been a strength, creating a rhythm that keeps stakeholders accountable and responsive to AI's rapid evolution. Regular summits have fostered timely policy development and raised awareness, with each meeting building purposefully on previous achievements. For example, in Seoul, states secured commitments from companies to present risk management frameworks in Paris. The regular cadence proved essential for translating initial discussions into firm commitments.

---

[3] Current and subsequent generations of general-purpose or specialized AI systems that match or exceed the capabilities of the most powerful available systems

[4] The Global Partnership on AI is an international initiative launched by France and Canada in 2018 seeking to support the responsible development of AI. It is now hosted by the OECD.

**Driving action.** The summits have motivated governments and industries to act. High-level engagement, including head-of-state involvement, has complemented other diplomatic efforts while encouraging corporate prioritization of AI safety. These gatherings have empowered internal safety teams at AI companies and driven cross-industry collaboration, resulting in concrete commitments and progress. This dual impact – within organizations and across the sector – has made the summits a unique driver of accountability and governance in AI. The commitments made at previous summits have advanced AI governance and provided a platform that other governance initiatives, such as the EU General-Purpose AI [Code of Practice](), have built upon.

Through these three mechanisms, the series has established itself as a leading platform for international discussions on AI governance, raising awareness about the speed of development and the risks of advanced AI and causing certain countries to prioritize AI governance in their political agendas. While primarily discussion-based, the summits have proven to be a practical mechanism for influencing AI governance towards the public interest. To maintain its momentum and impact, the summit series must continue to deliver concrete outcomes and adapt to the changing technological and geopolitical landscape.

As the series approaches its third summit in Paris, refining its structure and strategy will become essential to secure its long-term impact. While the current model has delivered significant achievements, it relies heavily on host nations' ability to quickly organize complex events and secure commitments. As more fora emerge for discussing AI governance, the series must also evolve to maintain its distinctive contribution. Furthermore, the early summits benefited from host countries with expertise in advanced AI safety and governance – a capability that future hosts may not share to the same degree. This memo thus examines key elements for strengthening the summit series to ensure that it remains an effective and sustainable forum for advancing AI governance.

## 02.    Key design considerations

As the summit series adapts to emerging technologies, a crowded governance ecosystem, and shifting political priorities, key design elements must be clearly defined. This section outlines considerations to ensure that the series remains strategically impactful, complements existing initiatives, and shapes the future of AI governance effectively. We focus on opportunities and tradeoffs in the series' relationship with other AI governance initiatives, scope, participation framework, and form of institutionalization. The analysis is informed by conversations with experts, workshops, other international convenings, as well as the successes and challenges of the series so far.

**2.1. Scope considerations**

Defining the scope – the range of issues to be addressed and the boundaries of work – is a critical design decision for any international initiative. A clear scope helps avoid duplication, attracting participants and facilitating progress. For instance, the [Montreal Protocol](#)'s[5] success in phasing out ozone-depleting substances stemmed partly from its clearly defined scope.

Closely related to scope is the mandate of the summit series – its authority and legitimacy to address certain issues and pursue specific goals. The mandate, whether implicit or explicitly defined through political agreements among participating countries, provides the necessary political backing for the series to be effective. A clear mandate helps define a clear scope, and vice versa. The mandate will be stronger if it is explicitly agreed upon among the participating countries, providing legitimacy.

Considering the overarching need for a clear scope and mandate, and the foundational premise that advanced AI governance must remain the central focus, this analysis explores three potential approaches to defining the scope of the summit series:

1) Governance of Advanced AI: This approach, building on the Bletchley Summit, would narrow the focus to the safety of advanced AI. While allowing for concentrated progress on critical risk-related issues, this scope may be perceived as overly focused on potential negatives, sidelining broader opportunities presented by these technologies.

2) Governance of Advanced AI, Plus Opportunities: This approach broadens the focus to include the potential benefits of advanced AI. It could address public interest issues such as AI's potential to accelerate scientific discovery and develop solutions to pressing global problems, as well as how to ensure equitable access and improve foresight regarding future developments and their societal impacts. This approach could attract a wider range of stakeholders and promote a balanced consideration of both risks and benefits, while maintaining a clear focus on advanced AI. Precedents for this exist in the Seoul Summit.

---

[5] The Montreal Protocol on Substances that Deplete the Ozone Layer (the Montreal Protocol) was an international agreement made in 1987. It was designed to stop the production and import of ozone-depleting substances and reduce their concentration in the atmosphere to help protect the earth's ozone layer.

3) Governance of Advanced AI, Plus Other Topics: This option expands the scope beyond advanced AI to include broader AI-related issues not specific to these systems. For instance, it might encompass general challenges in cybersecurity, manipulated content, and misinformation, rather than focusing specifically on how advanced AI might exacerbate these problems. This is the approach taken by the French AI Action Summit, which introduced multiple tracks on topics such as the future of work or innovation and culture, providing an interesting model for expanding participation and addressing diverse topics. This wider scope, while potentially more inclusive, raises questions about the appropriate participants and whether all these issues need coordinated action at this level. It could shift the focus from a core group of advanced AI developers to a much larger group of technology companies, potentially complicating the series' ability to foster international cooperation by entangling it in broader social and regulatory debates, like those surrounding social media governance.

To maintain a balance between broad engagement and focused outcomes, **we propose a streamlined two-track structure. The first and core track (Track 1) would focus on the governance and safety of advanced AI**, building on the mandate established in earlier summits and primarily involving countries with significant capabilities in advanced AI development, along with major AI companies. **The second track (Track 2), building on the French precedent, would explore the public interest opportunities presented by advanced AI,** allowing for broader participation and discussion.

This two-track structure allows for exploration of both the risks and benefits of advanced AI while maintaining a manageable scope. If political declarations are used to signify commitment, as in previous summits, it is recommended to focus political efforts on Track 1, helping maintain momentum on advanced AI governance. This does not preclude other meaningful deliverables from the second track.

A potential criticism of this approach is that focusing solely on advanced AI may also create artificial divisions within the broader field of AI governance. However, given the unique challenges and potential impact of advanced AI and the need to ensure coherence amid the proliferation of different levels of regulations across jurisdictions, we believe that a dedicated track for advanced AI remains justified.

> **Relationship of the summit series with other international AI governance initiatives**
>
> The international AI governance ecosystem includes established initiatives like those at the G7, G20, BRICS, GPAI-OECD, and the United Nations (UN). While these fora address various AI-related issues, from ethical guidelines to standards, they do not specifically focus on the unique challenges posed by the most capable AI systems, often referred to as "advanced AI." These systems present distinct risks and opportunities that require dedicated attention.
>
> The summit series can carve out a valuable niche by concentrating on advanced AI, complementing existing initiatives. For example, it can build upon the work of specialized initiatives like the International Network of AI Safety Institutes (AISI Network[6]), which focuses on AI safety, and address the broader governance challenges specific to these powerful systems with the group of countries participating in the summits.
>
> While maintaining this focus, the series should also regularly engage with more universal efforts, such as those outlined in the UN's Global Digital Compact, being a Global Dialogue on AI Governance. The summit series can act as an agile complement to this broader dialogue, spotlighting advanced AI issues and contributing insights to inform the larger conversation. Achieving this synergy requires ensuring complementarity between the series' focus on advanced AI and that of other fora, and building deliberate bridges between the series and other relevant processes – for example by aligning meeting schedules and sharing information. For example, exploring mechanisms for collaboration between the AISI Network and countries not currently participating in it could enrich the discussions and prevent fragmentation.

## 2.2. Participation framework

A well-designed participation framework is important for the summit series' legitimacy and effectiveness. "Participation" encompasses both who is invited to attend and who can contribute to shaping agreements. "Legitimacy," in this context, refers to the broadly recognized fairness, representativeness, and authority of the summit as a venue for decision-making on advanced AI governance. A core tension exists between the desire for broad, inclusive participation and the need for a focused, manageable group capable of reaching concrete outcomes:

1) One approach would be to strive for a global summit, mirroring the UN's universal membership. This would maximize inclusivity and ensure that decisions carry weight across the international community. However, such a large and diverse group could also lead to increased complexity, dilute the summit's focus, and hinder the ability to reach agreements, replicating challenges often faced by large-scale international institutions ([Victor, 2006](#)).

2) Alternatively, a more limited participation model would prioritize efficiency and the ability to reach concrete outcomes. This approach could involve a smaller group of countries with significant capabilities in advanced AI development and deployment. However, this raises the question of who participates and on what basis. Determining which countries – and companies – qualify as "AI-leading" is not straightforward. Criteria for participation need to be carefully considered to avoid perceptions of exclusivity or bias. Factors to consider could include demonstrated technological capabilities, established AI research infrastructure, experience in AI governance, and the need for geographical diversity.

---

[6] The AISI Network brings together the AI Safety Institutes and equivalents of 9 countries (Australia, Canada, France, Japan, Kenya, the Republic of Korea, Singapore, the United Kingdom, and the United States) and the European AI Office to collaborate on AI safety research and testing.

A robust participation framework also needs to determine the role of non-state actors. Industry leaders, technical experts, and civil society bring essential perspectives and capabilities. The French AI Action Summit hosted affiliated events and established open working groups during the period in between summits; this model offers one avenue for incorporating these voices.

Another consideration is the relationship with existing initiatives, such as the AISI Network. The participation framework should ensure complementarity and avoid creating a sense of exclusion for those not directly involved in these initiatives. For example, if the summit series were to rely heavily on the AISI Network, countries like China or other stakeholders not currently part of that network might feel excluded, potentially discouraging their engagement and undermining the summit's legitimacy. These interdependencies should inform the design of the participation framework.

Lastly, the summit series should actively incorporate perspectives from the Global South. These perspectives offer valuable insights into the diverse socio-economic and cultural impacts of AI, particularly regarding how advanced AI might affect existing global inequalities. This requires exploring mechanisms that not only facilitate engagement but also ensure that the outcomes of the summit series address the specific needs and concerns of the Global South.

**Considering these complexities, a two-track approach may offer an effective solution.** The first track would involve a core group of countries, along with major AI companies with an observer status, who would be responsible for driving discussions and forging agreements on the governance of advanced AI (Track 1). This model recognizes that a smaller group can reach actionable outcomes more effectively while ensuring that the summit series does not duplicate existing decision-making spaces, where countries with fewer resources are already engaged in global institutions addressing broader AI-related issues. The second track would allow for broader participation in discussions related to the public interest opportunities presented by advanced AI (Track 2). This approach balances the need for focused, outcome-driven discussions with the importance of inclusivity and diverse perspectives.

For Track 1 membership, criteria for participation must be established and reviewed periodically. Criteria could include, but need not be limited to, national AI capabilities, jurisdiction over entities developing state-of-the-art AI, concentration of expert talent, and/or established or developing regulatory frameworks. Based on these criteria, Track 1 would potentially include Australia, Canada, China, the European Union, France, Germany, India, Italy, Japan, Netherlands, Singapore, Spain, South Korea, the United Arab Emirates, the United Kingdom, and the United States.

## 2.3. Institutionalization for continuity

In the context of international summits, "institutionalization" refers to the process of establishing formal structures, rules, and procedures that govern how the series operates. "Continuity" is about maintaining consistent progress and engagement over time. Both are important considerations for the long-term effectiveness of the summit series.

The primary challenge facing the summit series is not simply choosing between a flexible or rigid structure, but securing the sustained political will and buy-in needed to establish and empower any form of lasting institution. The series' achievements so far have been driven by the commitment of participating countries and stakeholders. Maintaining this commitment, while navigating the diverse interests and priorities of participants, will be more important than any specific institutional design. The current flexible approach – where the summits were organised without formal institutional structures – encouraged this initial buy-in.

However, for the series to achieve long-term impact, some degree of institutionalization may be necessary. More formal structures – such as a permanent secretariat, defined membership rules, and standardized procedures – could enhance consistency across summits, reduce administrative burdens (particularly for countries with limited resources), and help maintain momentum. For example, a secretariat could provide logistical support, manage communications, and track commitments, ensuring follow-up between summits. Defined membership rules, as discussed in the previous section, could clarify expectations and streamline decision-making.

However, establishing such structures requires a willingness among participants to cede some degree of autonomy and invest in a more formalized process. **This is where the core tension lies: finding the level of institutionalization that will strengthen the series' effectiveness without undermining the very political will that underpins it.** Too much rigidity could limit a host nation's ability to tailor an event to its national interests, industry needs, and political priorities, or adapt to emerging technological developments. It could also deter participation if countries feel overly constrained by formalized rules. The current, more flexible approach has secured high-level political commitment and ensured strong engagement by allowing each host to shape the summit's focus however these could also be attributed to the determined efforts made by the Heads of State from the previous hosting countries (eg. U.K and France)

High-profile host involvement elevates the summit's prestige and diplomatic importance, enabling "sherpa-level" negotiators[7] to leverage the summit's visibility to secure meaningful commitments from participants. This approach, however, can result in inconsistencies across summits – both in format and outcomes. And it also raises questions about how to maintain continuity. Increased reliance on institutionalized structures could address these challenges. Yet this same standardization risks diluting the high-profile political engagement that comes from host nations adjusting the summit to fit their own agenda.

The previous two summits were successful, and the flexible model has worked well so far. However, some degree of institutionalization could help ensure continuity, improve coordination, and support the long-term impact of the series. While flexibility is important for innovation, adding structure could make it easier to address challenges effectively over time. This could involve formalizing an organizational backbone to provide continuity, institutional memory, and operational support – for example, through a dedicated secretariat. A tiered participation model, as previously discussed, with a more formalized approach to the governance of advanced AI track (Track 1), could provide the right mix of structure and flexibility, while maintaining the necessary buy-in and preserving the adaptability to technological change. The degree of formalization should be calibrated, building upon the successes of the series to date and, most importantly, driven by the continued political will of participating countries. The next section explores specific design options, including the level of institutionalization, that could further strengthen the summit series.

---

[7] A sherpa is the personal representative of a head of state or head of government who prepares an international summit

## 03. Core design elements

The long-term effectiveness of the AI summit series depends on structural decisions about its operation. While the previous section explored broad considerations that should guide the summit series' development – its scope, participation model, and level of institutionalization – this section focuses on specific operational decisions that must be made. Below, we outline concrete design elements and their implementation options, drawing from existing international frameworks and governance models. These include host arrangement, secretariat format, participant selection processes, agenda-setting mechanisms, and meeting frequency. For each element, we analyze various approaches and their implications for the series' effectiveness.

**Table 2. Core design elements for the AI summit series**

| Design feature | Options |
|---|---|
| 1. Hosting arrangement | **Rotation of hosting** - Hosting passes between core members on a fixed schedule. |
| | **Bidding for hosting** - Members vote for candidates. |
| | **Regional groups hosting** - Hosting rotates between regions. |
| | **Joint hosting** - Two or more countries share hosting duties. |
| 2. Secretariat format | **Light coordination** - Host handles with a small working group. |
| | **Hybrid/incremental** - Core team retained, host has leeway. |
| | **Formal secretariat** - Fully independent from host, hosted by intl. org. |
| 3. Participants selection | **Host-driven selection** - Host country decides invitations. |
| | **Curated membership** - Predetermined participation criteria. |
| | **Blended model** - Core group + host-invited guests. |
| | **Universal participation** - Open to all countries. |
| 4. Agenda setting | **Host-led** - Host country sets primary agenda. |
| | **Multiple workstreams** - Parallel tracks with different goals. |
| | **Coordinated succession** - Multiple hosts coordinate over an extended period. |
| | **Steering committee** - International. Multi-stakeholder group sets priorities. |
| 5. Summit Frequency | **Annual summits with ongoing engagement.** |
| | **Flexible biannual summits.** |
| | **Annual summits with interim meetings.** |

### 3.1. Hosting arrangement

The current system for selecting summit hosts relies on an ad hoc process. This can introduce uncertainty about future summits and contribute to variations in planning and engagement. Without a predictable hosting mechanism, participants may find it difficult to coordinate efforts or develop long-term strategies across multiple events. Host nations also differ in their available resources and networks, which can affect the overall quality and consistency of each summit.

Host selection involves answering several questions: Should hosts be chosen based on geographic considerations? Should preference be given to countries with significant AI-producing capabilities or demonstrated leadership in AI governance? Should financial or logistical readiness be considered? And overall, how does the selection process work?

From a procedural perspective, a clear, agreed-upon method for selecting hosts can improve fairness, transparency, and alignment with the summit's goals. Many informal international fora – such as the G7, G20, or GPAI – use structured host selection procedures that may offer models for a more standardized approach.

Another important factor is the selection process's impact on participation. To maintain the summit's effectiveness as a forum for international cooperation, host selection should favour neutral venues. This approach helps ensure participation from major AI powers, whose engagement is essential for meaningful progress on AI governance.

Once selected, the **host's authority over the summit** is another critical aspect, as well as their control over the agenda. We briefly covered some aspects of this in Section 2.2 and offered a proposal to resolve the tensions related to participant choice. Participation rules determine who attends, while content-related decisions guide what is discussed or decided. If a single host sets the agenda with little input from others, the summit may achieve decisive outcomes but risk looking only after the host's national interests. If agenda-setting is highly open, the event may include a broad range of viewpoints but could face challenges in maintaining focus. Agenda-setting will be covered later in this section.

Procedures for host selection could draw on several models to establish a more standardized approach. We outline four potential models, all of which **centre on a core group of AI-leading countries (track 1) managing hosting decisions.** As the field evolves and new actors emerge, both the composition of this group and the suitability of these models may require periodic reassessment.

Option A: Full rotation

Under a straightforward rotation system, hosting duties pass between core participating countries on a fixed schedule. This approach is similar to the G7 model, where the rotation order is pre-established. The summits could rotate between individual countries or between regional groups to alternate regions. This approach creates predictability and offers each core country an opportunity to shape the series. The host would manage logistics and set the agenda in consultation with previous hosts, maintaining a degree of continuity. However, this approach may not always align with the summit's thematic needs or a country's readiness to host.

Option B: Bidding system

A bidding system presents another possibility, where countries propose their vision for hosting and members vote for candidates. This approach draws from the COP hosting process. It could help ensure hosts have adequate resources and commitment. However, this system might favour well-resourced countries and create political tensions around selection.

Option C: Regional groups

Instead of rotating among individual countries, hosting could rotate between geographic regions or groupings (e.g., Africa, Asia-Pacific, Europe). This may enhance global representation and distribute responsibilities more evenly, though it could face challenges in regions with varying capabilities and geo-economic interests. UN conferences like COP and G20 follow this approach.

Option D: Joint hosting

Two or more countries share hosting duties, potentially pairing nations with differing resources or priorities. This model can distribute costs, foster collaboration (such as between Global North and Global South), and ensure continuity, yet may introduce higher coordination demands and risk stalemate if co-hosts disagree on key issues. An example of this might be the Seoul Summit, where the UK and South Korea worked together to organize the summit.

To illustrate various hosting models, Table 2 summarizes four possible approaches (A–D), each with its own advantages and drawbacks.

**Table 3. Hosting arrangement options.**

| Model | Description | Pros | Cons | Example |
|---|---|---|---|---|
| A: Rotation of hosting | Hosting duties pass between core participating countries on a fixed schedule, potentially with regional alternation. | Predictable schedule. Equal hosting opportunities. Fosters continuity through host collaboration on logistics and agenda. | Summit quality may vary depending on host resources and experience. May not always align with thematic needs. | G7, G20 (with regional alternation) |
| B: Bidding for hosting | Countries propose their vision for hosting; members of a core group vote for candidates. | Competition drives innovation. Allows hosts to showcase their strengths. Ensures hosts are adequately resourced and committed. | Might favour well-resourced countries. Can create political tensions. May lead to inconsistencies in summit focus. | COP hosting process |
| C: Regional groups hosting | Hosting rotates between geographic regions or established groupings (for example, Africa, Asia-Pacific, Europe). | Enhances global representation. Distributes responsibilities more evenly. Leverages existing regional cooperation frameworks. | Regions with varying capacities and interests may face challenges. Can lead to inconsistencies in summit quality and focus. | UN conferences, G20, COP |
| D: Joint hosting | Two or more countries share hosting duties, potentially pairing nations with differing resources or priorities (for example, Global North and Global South). | Distributes costs. Fosters collaboration and knowledge sharing. Bridges different perspectives. Provides built-in continuity. | Higher coordination demands. Risk of deadlock if co-hosts disagree. May be more complex to organize. | Seoul Summit (UK and South Korea co-hosted) |

The selection of summit hosts must take into account considerations related to procedural clarity and fairness, geographic diversity, adequate host capabilities, and participation from major AI powers. It's equally important to provide advance notice to enable thorough preparation.

**We propose a hybrid approach combining options B and C: the summit would rotate between regions, but within each region local Track 1 countries could bid to host.** The decisions would be made two years in advance, and a steering group will be created to work together with past, present and future hosts to ensure continuity. The two years of advance notice would provide preparation time. The structured nature of the process would reduce uncertainty for all stakeholders.

**3.2. Secretariat format**

A secretariat can serve as the administrative backbone of a summit series by managing logistics, coordinating agendas, and facilitating communication among participants[8]. At the same time, some summits operate with no secretariat at all, relying instead on ad hoc arrangements. Between these extremes lie various "lighter" models, such as those employed by the G7, where a small administrative support team exists, yet the host retains considerable flexibility. It seems plausible that the larger the number of members, the more difficult it is to keep an effective and yet informal support system in an outcome-driven event like the summit series.

The role of a secretariat, broadly understood, is to provide support for the coordination and continuity of a summit series. However, a fully formalized secretariat is not always necessary, especially in the early stages of an initiative. Some international efforts have relied on informal, adaptable arrangements that evolve organically over time based on demonstrated needs. Moving too quickly toward a formal secretariat can also be resource-intensive and may diminish the series' flexibility. A more gradual approach allows participants to build trust, test the summit's value, and identify whether or what type of secretariat is needed.

Deciding whether and how to establish a secretariat involves finding the right balance between structure to ensure continuity and enough political buy-in from all parties to support such an effort.

Below is a **progressive framework for thinking about secretariat options, beginning with minimal oversight and building toward full institutionalization**. The final choice depends on how the summit series weighs trade-offs such as resource availability, host flexibility, and the need for long-term consistency.

*Option A: Light coordination (e.g., Troika model, G7-style, host-led)*

Under a "light coordination" system, the summit host handles most responsibilities, possibly aided by a small working group. The G20's troika format – where past, current, and future hosts collaborate – facilitates knowledge transfer without creating a permanent institution. The G7 uses a lean administrative team that changes with each presidency, preserving enough continuity through informal networks. This model enables agile responses to changing political priorities but can result in irregular record-keeping, varying quality, and potential knowledge loss.

---

[8] A secretariat is a permanent administrative office or department that handles the day-to-day operations and organizational tasks of an international organization or summit series.

*Option B: Hybrid/incremental*

A semi-permanent core team is retained for multiple summits, ensuring modest institutional memory and stable operations, while the host still has leeway in shaping agendas and inviting participants. Rotating staff or partial rotation (e.g., a small permanent unit plus a host-appointed team) can maintain a middle ground. Over time, if the summit's scope widens or membership increases, this model can scale up its administrative resources and official mandates without immediately becoming a full-blown bureaucracy. An example could be the role the OECD plays for the G20, although this is not exempt of controversy[9].

*Option C: Formal secretariat*

A fully independent[10], permanent secretariat – often underwritten by an international organization or its own agreement/treaty-based structure – provides the strongest continuity, coordination, and specialized expertise (for example, OECD for GPAI, UN Secretariat or ASEAN Secretariat). It also offers neutrality and a centralized authority to track commitments and follow-through between summits. A dedicated secretariat can sometimes help to prevent influence by powerful stakeholders, but it may pose a risk of influence by the host of the secretariat. Creating such a structure typically involves negotiations and funding, and it may hinder trust, depending on who is part of the international organization hosting the secretariat (e.g. China is not part of the OECD).

The table below combines the progressive approach with the more detailed categories sometimes discussed in policy circles (Independent Secretariat, Semi-Permanent Hybrid, Rotating Secretariat, Host-Controlled). In practice, these categories can map onto the three overarching levels of institutionalization (Light Coordination, Hybrid, Formal Secretariat).

---

[9] Wouters, Jan, & Van Kerckhoven, Sven. (2011). The OECD and the G20: an Ever Closer Relationship. *George Washington International Law Review, 43(2),* 345-374.

[10] In this context, "independent" means that the secretariat operates autonomously from any single host government or stakeholder. It typically has its own funding mechanism, staffing, and decision-making structures, ensuring that no single participant – be it a country, company, or institution – exercises exclusive control. An "independent" secretariat might be hosted by a neutral international organization or formally established under an agreement-based framework, so it can maintain continuity, neutrality, and professional management regardless of which host or coalition of hosts is leading the summit series at any given time.

**Table 4. Secretariat format options**

| Option | Description | Pros | Cons | Examples |
|---|---|---|---|---|
| A: Light coordination<br><br>Low level of institutionalization | The host country manages most logistics and coordination, possibly with a small support team or working group. | Agile and responsive to changing priorities.<br><br>Lower cost.<br><br>Host retains flexibility. | Inconsistent record-keeping.<br><br>Potential for knowledge loss.<br><br>Variable quality of support.<br><br>More difficult with large number of members. | G7, G20 (Troika model) |
| B: Hybrid/ incremental<br><br>Medium level of institutionalization | A core team provides continuity across multiple summits, while the host still has significant influence on the agenda and overall direction. | Modest institutional memory.<br><br>Stable operations.<br><br>Scalable to meet evolving needs.<br><br>Balances continuity with host flexibility. | Requires some ongoing funding and coordination.<br><br>May be less responsive than a fully host-controlled model. | OECD's role for the G20 (though controversial) |
| C: Formal secretariat<br><br>High level of institutionalization | A permanent, independent body with dedicated staff and resources provides administrative and logistical support. | Strongest continuity and coordination.<br><br>Specialized expertise.<br><br>Neutrality and central authority.<br><br>Can help to prevent undue influence by powerful stakeholders. | Resource-intensive.<br><br>Requires complex negotiations and funding.<br><br>May reduce flexibility.<br><br>Risk of influence by the host of the secretariat.<br><br>Trust can be hindered depending on the host organization. | OECD for GPAI, UN Secretariat, ASEAN Secretariat |

In most international contexts, countries first create a formal organization through a treaty or intergovernmental agreement, and that organization's secretariat then convenes summits as part of its mandate. Examples include the UNFCCC Secretariat, established by the United Nations Framework Convention on Climate Change, which organizes the annual COP meetings, or the OECD Secretariat. In these "classic" arrangements, the summits are just one product of a pre-existing institution whose mission and governance rules are set out in founding documents. By contrast, the summit series discussed here preceded any formal institution or initiative.

Rather than being a formal organization from the start (like GPAI was), it has been operating with an ad hoc model. Moving toward institutionalization, with a permanent secretariat would therefore be a "reverse" path to institutionalization, where a formal structure grows out of an existing summit process rather than the other way around. This reverse path offers flexibility and lowers initial overhead, but it may require more complex negotiations later if countries decide they need a framework and a secretariat to manage an expanding range of issues and participants.

### 3.3. Participant selection

As pointed out earlier, every summit grapples with the same fundamental questions: Who gets a seat at the table? On what basis? And who makes that decision? This summit series adds another layer of complexity by incorporating non-state actors, a challenge that seems somewhat easier since these non-state actors are the leading companies and labs working on advanced AI and those are only a handful.

In the brief history of this series, the host has typically held the prerogative to determine participation. For instance, the UK government invited 28 participants in total to Day 1 of the Bletchley Summit at Ministerial level, covering Australia, Brazil, Canada, Chile, China, the European Union, France, Germany, India, Indonesia, Ireland, Israel, Italy, Japan, Kenya, the Kingdom of Saudi Arabia, the Netherlands, Nigeria, the Philippines, the Republic of Korea, Rwanda, Singapore, Spain, Switzerland, Türkiye, Ukraine, the United Arab Emirates, and the United States of America. This is the group that signed the Bletchley Declaration that was published on Day 1 of the Summit on 1 November 2023. For the leaders' session on Day 2 of the Bletchley Summit a subset of these countries participated. This effectively created a two-pronged approach. All 28 participants attended, engaged in discussions, and signed the Bletchley Declaration commitments, while a smaller group at Head of State, Head of Government/Leader level agreed on an additional statement.[11]

The subsequent Seoul Summit, co-organized by the Republic of Korea and the United Kingdom, brought together the same 28 participants in total and had the same format as Bletchley. There was a Ministerial-level day involving all 28 participants and a virtual meeting with the same smaller leaders' group as at Day 2 of Bletchley. The key outcome of the Seoul Summit was the agreement by companies from North America, Europe, the Middle East, and Asia (including China) to the Frontier AI Safety Commitments. This was the first time that there had been a genuinely global agreement involving such a geographically wide range of advanced AI companies. Finally, the upcoming AI Action Summit, organized by the French government, is expected to host more than 90 countries – a notable shift toward wider international engagement.

---

[11] Safety Testing: Statement of Session Outcomes, 2 November 2023

These differences in the number of participants and the level of their involvement highlight the trade-offs inherent in different participation models. By "level of involvement," we mean the extent to which countries can actively shape the agenda, participate in negotiations, influence the final outcome, and draft or sign official documents. This can range from simply attending and observing, to actively contributing to discussions, to having a direct hand in negotiating and endorsing final agreements. Additionally, the main focus of each summit has largely followed the host country's national priorities, rather than sticking to a consistent theme across the series.

Before we explore potential models for choosing participants, it's helpful to briefly look at the main options: At one end of the spectrum is open participation, similar to UN processes. This model offers wide representation and inclusivity, avoiding the criticism often directed at exclusive groups ("a small club of rich countries making decisions for the rest of the world"). However, while this approach scores high on legitimacy, it can become difficult to manage, potentially hindering the summit's ability to achieve concrete results, as often seen in larger international forums.

On the other end of the spectrum is a select group of participants, a model often used by coalitions of "like-minded countries" (such as the G7, GPAI, and OECD) or groups united by a common goal (like ASEAN, G20, and NATO). In the context of advanced AI, one could argue for **a new informal group made up of the most advanced AI nations.** This would primarily involve countries with proven expertise in AI technologies and their governance. This approach could make decision-making more efficient and boost the summit's ability to deliver tangible outcomes. However, this would likely come at the expense of perceived legitimacy and could face initial criticism due to its selective nature.

This brings us to the main question: What are the options for deciding who attends these summits, and what are the pros and cons of each? One option is to stick with the current system, where the host nation decides. A variation could involve a "troika" system, where the current host along with the previous and next hosts jointly determine the guest list. This would offer some predictability and continuity, giving the organizers of the next summit a 6-month to 1-year window to prepare, depending on how often the summits are held.

However, other models could involve a core group of leading AI nations taking a more active role in shaping the summit's future. These nations would become the main decision-makers when it comes to adding new participants and attendees.

Below we look at four distinct options for participant selection, each with its own set of advantages and disadvantages:

**Table 5. Participant selection model options**

| Option | Description | Pros | Cons |
|---|---|---|---|
| A: Host-driven selection | Host nation decides participants, potentially consulting with a troika (past, current, future hosts). | Keeps the host in control. Allows for tailored agendas. Potentially faster decisions. | Lack of continuity. Potential for inconsistency in who attends. Risk of perceived favouritism. |
| B: Curated membership | A set group of nations with membership based on specific criteria (e.g., AI development, regulatory frameworks, FLOP?). | More continuity. Focused discussions. Potentially stronger commitment from members. A clear path for others to join. | Potential for a few powerful nations to dominate. Difficulty in defining and agreeing on fair criteria. |
| C: Blended model | A core group of permanent members, plus a number of guest countries chosen by the host. | Balances continuity with flexibility. Allows for diverse perspectives while maintaining a core group of committed countries. | Potential for tension between core members and guests. |
| D: Universal participation | Open to all countries, potentially with mechanisms to ensure regional representation. | Maximum inclusivity. Broad legitimacy. Diverse perspectives. | Difficult to make decisions. Risk of watered-down outcomes. Potential for the agenda to become too broad. |

Option A: Host-driven selection:

This model gives the host maximum flexibility to shape the guest list. While it allows for agility and responsiveness to emerging priorities, it may lack continuity and could be perceived as less legitimate if the selection process seems opaque or biased. A "troika" consultation could address some of these concerns by incorporating input from past and future hosts, fostering a degree of shared ownership.

Option B: Curated membership:

This approach prioritizes continuity and in-depth engagement by establishing a set group of countries. Membership would be based on pre-agreed criteria, ensuring that participating nations have a proven commitment to and capability in AI governance. This model could foster more focused discussions and potentially lead to stronger commitments. However, it risks excluding voices and could be perceived as elitist.

Option C: Blended model:

This model tries to balance the benefits of continuity and flexibility. A core group of permanent members would provide stability and institutional memory, while the host's ability to invite guest countries would allow for the inclusion of diverse perspectives and the ability to address new issues or engage with countries making significant progress in AI.

Option D: Universal participation:

This model prioritizes inclusivity and broad legitimacy by opening the summit to all interested countries. Mechanisms for regional representation could be included to ensure a balanced geographical spread. While this approach maximizes the diversity of voices, it could lead to unwieldy decision-making and potentially result in less ambitious outcomes due to the need to accommodate a wider range of interests and capabilities.

Choosing the best participation model depends on the desired balance between inclusivity, effectiveness, and the long-term goals of the AI summit series. It will shape how effectively the summit series addresses its **dual mandate: fostering advanced AI governance among key actors while incorporating broader global perspectives**. Smaller groups (e.g., Options A or B) allow for focused, outcome-driven decision-making but may attract criticism for being exclusionary. Broader groups (e.g., Option D) enhance legitimacy but risk operational inefficiencies or diluted focus. By carefully balancing these factors, the summit series can refine its participation framework to ensure that major AI powers make meaningful commitments, while other regions and voices contribute constructively to the dialogue. This approach would allow the summit to maintain credibility, address advanced AI challenges, and evolve as a globally relevant platform for AI governance.

A structured, transparent process for selecting participating countries might enhance the series' credibility. Beyond state actors, the participation of AI companies and other non-state stakeholders presents a critical choice. Including major AI companies could enhance technical relevance and implementation capability while securing private sector buy-in for summit outcomes. Conversely, excluding private actors affirms a stronger role for states but potentially reduces the practical impact and technical depth of discussions.

An additional consideration is the level of government participation expected at these summits. Current variations in leadership attendance – with some countries represented by heads of state and others by lower-level officials – can create tensions and signal inconsistent commitment levels. The G20 and APEC summits provide useful models, with clear expectations for head-of-state participation and delegation composition.

## 3.4. Agenda setting

The agenda of a summit functions as its roadmap, influencing discussions, driving outcomes, and determining its lasting impact. A thoughtfully constructed agenda ensures that the series addresses pressing issues, builds on past progress, and fosters engagement. Currently, the host sets the agenda, allowing each host to align the summit's focus with its own priorities. While this offers clear leadership, it may also hinder continuity between meetings and potentially lead to a fragmented approach to complex, long-term challenges like the governance of advanced AI. Therefore, a key decision for the evolution of this series is to clarify how, when, and by whom the agenda is determined.

This section explores various agenda-setting models, evaluating their effectiveness in balancing host flexibility with sustained progress on core issues.

Option A: Host-led

This model retains the current system, where each host country determines the primary discussion topics and deliverables for its summit. The host nation typically collaborates with technical experts and other stakeholders to develop a detailed agenda but retains final decision-making authority. This approach offers the advantage of clear ownership, as the host has a strong incentive to ensure the summit's success. It also provides flexibility, allowing each host to address its most pressing concerns and showcase leadership in specific areas. However, this model can lead to a fragmented approach, with topics shifting significantly from year to year. The quality and relevance of the agenda may vary depending on the host's capacity and priorities, and important ongoing initiatives may be neglected if not prioritized.

Option B: Multiple workstreams

Similar to the G20 structure or the French approach, this model establishes parallel workstreams alongside the main government track. These groups would develop their own agendas and recommendations within their respective areas of expertise, feeding into the main summit discussions. This approach provides a platform for diverse voices, leverages specialized knowledge, and allows for continuous work on specific issues, even as the main summit's focus shifts. However, managing multiple workstreams and ensuring coherence between them can be complex. Parallel workstreams may operate in silos, hindering integrated approaches to complex problems, and the model requires significant resources and coordination.

Option C: Coordinated succession

This model draws inspiration from the European Council's presidency trio system, where three successive host countries collaborate to develop a shared 18-month agenda framework. Each host retains autonomy over the specific priorities and deliverables for its individual summit but works within the framework of an agreed-upon roadmap outlining core agenda items, priorities, and desired outcomes. This method aims to ensure smoother transitions and medium-term continuity while preserving host country ownership. This approach facilitates a more coherent and sustained approach to key issues across multiple summits and offers greater predictability about the direction of the series over the medium term, which makes it easier for stakeholders to anticipate and engage with the agenda. On the other hand, it requires close collaboration and agreement among the three hosts. Larger or more influential countries within the trio could have disproportionate influence, and the ability of individual hosts to respond to new developments may be somewhat limited. The roadmap should thus undergo periodic review and updates to remain relevant to emerging challenges and technological advancements.

Option D: Steering committee

In this option, an international steering committee would be established to shape the strategic priorities and ensure agenda coherence across summits. This steering committee could include representatives from Track 1 members  While the host country would maintain significant influence over the specific focus areas of its summit, the steering committee would provide guidance, ensure alignment with long-term objectives, and maintain progress on core issues. This approach, used in the past by the Global Partnership on AI, can be ineffective if the non-governmental stakeholders are merely token participants, lacking real influence in decision-making processes or the ability to shape meaningful outcomes. If taken seriously, this approach promotes continuity and a sustained focus on key priorities, leveraging diverse expertise to inform agenda development and ensuring that individual summits contribute to a broader, long-term vision. However, reaching consensus among diverse members could be challenging. The committee's composition and mandate would need careful consideration to avoid undue influence by any particular stakeholder group, and ensuring fair and balanced representation from different regions and sectors could be difficult.

These models are not mutually exclusive. Success will likely require combining elements from different models to create a hybrid structure that maintains the series' forward motion while adapting to changing circumstances and priorities. For example, a multi-year roadmap could be developed and overseen by a steering committee, providing both long-term direction and expert guidance. Alternatively, a coordinated succession model could incorporate multi-stakeholder workstreams, ensuring both continuity and inclusivity.

The choice between these models, or a combination thereof, should reflect both practical considerations of resources and capacity and strategic goals for the series' evolution. The optimal approach may also evolve, with the series potentially starting with a lighter-touch model and gradually moving towards a more structured approach as it matures and its scope expands. A commitment to ongoing evaluation and adaptation will be essential to ensure that the chosen model remains fit for purpose and capable of driving meaningful progress on critical issues.

**Table 6: Agenda-setting options**

| Model | Description | Pros | Cons |
|---|---|---|---|
| A: Host-led | Each host country determines the primary discussion topics and deliverables for their respective summit. | Clear leadership and ownership. Flexibility for hosts to address their priorities. | Limited continuity between summits. Variable agenda quality. Potential for oversight of important ongoing initiatives. |
| B: Multiple workstreams | Parallel workstreams managed by different stakeholders (e.g., business, civil society) alongside the main government track. | Inclusivity of diverse voices. Leverages specialized expertise. Sustained progress on specific issues. | Coordination challenges between tracks. Potential for fragmentation. Resource-intensive. |
| C: Coordinated succession | Three successive host countries collaborate to develop a shared 18-month agenda framework. | Enhanced continuity across summits. Balanced ownership among hosts. Predictability over the medium term. | Complex coordination required. Potential for domination by larger countries. Reduced flexibility for individual hosts. |
| D: Steering committee | An international multi-stakeholder group shapes strategic priorities and ensures agenda coherence across summits. | Strong institutional memory. Expert-driven agenda development. Strategic coherence across summits. | Potential for political disagreements. Risk of capture by specific interests. Challenges in ensuring fair representation. |

**We propose a multi-year roadmap with a Track 1 member steering committee.** This committee will regularly update the roadmap with expert input to guide the summit series' direction. The steering committee will set the Track 1 agenda, focusing on advanced AI governance. The host will set the Track 2 agenda, addressing broader public interest issues.

While the roadmap provides overarching guidance, hosts retain the ability to shape their individual summit agendas, highlighting specific priorities or addressing emerging issues within the established framework. This balance enables hosts to demonstrate leadership and address national concerns while contributing to the series' broader objectives. The model's success depends on a well-organized steering committee and a clearly defined process for developing the roadmap and integrating host input. It strikes an ideal balance between providing structure and allowing flexibility.

### 3.5. Summit frequency

The frequency of summits shapes their effectiveness. How often these meetings occur impacts the forum's ability to respond to rapidly evolving AI challenges, the depth of preparation participants can undertake, and the overall momentum of the initiative.

It's important to clarify that this AI summit series is not a formal governance institution with decision-making authority, like the UN or even the G20 in its capacity to influence global economic policy. Instead, it serves as a forum for dialogue, cooperation, and consensus-building among key stakeholders in the field of AI. The goal is to foster a shared understanding of the challenges and opportunities presented by advanced AI, explore potential areas for collaboration, and potentially facilitate the development of norms, standards, and best practices.

Experience in other international fora suggests that regular, predictable meetings can enhance commitment. This is especially true when meetings occur frequently enough to build upon previous outcomes. However, more frequent meetings also place greater demands on institutional support, potentially requiring a shift from the current ad-hoc hosting system to a semi-structured secretariat. Frequent meetings can create barriers to participation, particularly for less resourced countries. The time between summits must be sufficient for meaningful intersessional work – implementing previous commitments and preparing new initiatives – and the schedule should align with other international processes to avoid conflicts.

The current practice of holding AI summits every six months (biannually) represents a deliberate trade-off between responsiveness and capacity. While this allows for relatively agile decision-making and sustained high-level attention, it can strain institutional support mechanisms, especially without a permanent secretariat or standardized processes for work between summits. Without enough preparation time and consistent administrative support, there's a risk that these meetings could become repetitive, substanceless or with limited progress on implementing prior commitments.

Several options exist for determining summit frequency, each with its own set of trade-offs:

Option A: Annual summits with ongoing engagement

This model would involve a major annual summit, providing a focal point for high-level discussions, announcements, and agenda-setting for the year ahead. The annual summit would be complemented by ongoing engagement mechanisms, including regular virtual meetings (e.g., quarterly), ad-hoc working groups on specific issues, etc.

Option B: Flexible biannual summits

This model is based on the current practice of holding summits roughly every six months but with a more explicit acknowledgment of flexibility. The six-month interval would serve as a baseline, but the organizers, in consultation with key stakeholders, could adjust the timing of summits.

Option C: Annual summits with interim meetings

This model features a main annual summit for high-level discussions, complemented by an interim meeting that brings together the core group of countries for technical discussions. This interim meeting would focus on reviewing technical progress, discussing the interim "State of the Science" report, and preparing input for the main annual summit. The location of these interim meetings would be determined flexibly, with the host of the annual summit having the option to hold them either in their country, online, or elsewhere. This approach balances the need for broad stakeholder engagement with the need for more focused, technical dialogue among leading AI nations.

Each option has its own set of advantages and disadvantages, as shown in table 7:

**Table 7: Summit Frequency Options**

| Option | Description | Pros | Cons |
| --- | --- | --- | --- |
| A: Annual summits with ongoing engagement | Main annual summit for high-level discussions and agenda-setting. Ad-hoc working groups on specific issues. | Provides a predictable rhythm. Allows for in-depth work between summits. Offers flexibility through virtual meetings and working groups. Lower organizational burden. | May not be responsive enough to very rapid developments. Requires active participation in intersessional activities to maintain momentum. |
| B: Flexible biannual summits | Baseline of summits every six months. Organizers can adjust timing. | More frequent high-level interaction than annual summits. Offers flexibility to adapt to the evolving AI landscape. Allows summits to adjust to specific issues. | Could lead to uncertainty if the schedule shifts often. Requires careful coordination. May still place a significant burden on organizers and participants. |
| C: Annual summits with interim meetings | Main annual summit for high-level discussions. Interim meeting (in person) with a core group to discuss progress and a "State of the Science" report. | Balances high-level engagement with in-depth technical discussions. Provides a forum for presenting and discussing scientific advancements. Regular engagement from core group of countries. | Potential for a two-tiered system. Requires clear criteria for participation in the interim meeting. Adds complexity in planning and coordination. |

## 04. Conclusion

The AI Summit series has quickly become a significant forum for international cooperation on the governance of advanced AI systems. Its contributions include the first international declaration on advanced AI safety at the Bletchley Summit, securing voluntary safety commitments from leading AI companies across multiple continents, commissioning the International Scientific Report on the Safety of Advanced AI, and inspiring the creation of national AI Safety Institutes. By focusing on the specific risks and opportunities presented by the most capable AI systems, the series complements broader discussions occurring in other international bodies such as the G7, G20, OECD, and UN. The summits have demonstrated a capacity to drive action from both governments and industry. They have acted as a forcing function, pushing companies to prioritize frontier AI safety and prompting nations to implement mechanisms for better AI development and deployment. The regular cadence, even without formal structure, has fostered accountability and responsiveness to the fast-paced evolution of AI.

However, as the series progresses toward its third iteration in Paris (the AI Action Summit) and beyond, it faces the challenge of sustaining momentum and impact in an increasingly crowded landscape of AI governance initiatives. To maintain its relevance and effectiveness, the series must continue to refine its approach, guided by the key considerations and design elements outlined in this analysis. The summit series is more likely to continue delivering concrete outcomes if it maintains a clear and distinct niche, particularly in relation to other international AI governance efforts; adapts its scope to balance focus with inclusivity; develops a participation framework that ensures both representativeness and efficiency; and explores appropriate levels of institutionalization to provide continuity without sacrificing flexibility.

A central recommendation of this analysis is for the summit series to **retain its core focus on the governance of advanced AI**, either as a unique theme or through a core track. This focus addresses a gap in the current ecosystem, where no other forum is dedicated specifically to these most capable AI systems. Within this focus, a **two-track structure may offer an optimal balance. The first track would concentrate on the governance of advanced AI, primarily involving countries with significant AI capabilities and the major companies operating within their jurisdictions. The second track would explore the broader public interest opportunities** presented by these technologies, allowing for wider participation and a more diverse range of perspectives.

The participation model must also evolve to balance the need for efficiency with the importance of inclusivity. While a smaller group of AI-leading nations and companies may be necessary to drive concrete agreements on advanced AI governance, it is also essential to incorporate broader perspectives, particularly from the Global South, to ensure legitimacy and address potential global inequalities. Here again the two-track framework helps resolve this tension, with Track 1 responsible for decision-making on advanced AI, whereas wider Track 2 attendance can foster discussions on public interest opportunities led by the host country.

Regarding institutionalization, the summit series must find a middle ground between structure and flexibility. While the current ad-hoc model has enabled agility and high-level engagement, some degree of formalization is worth considering. This analysis explored several options, including: more structured hosting arrangements (e.g., rotation, bidding, regional groups, or joint hosting), a potential secretariat function (ranging from light coordination to a more formal, independent secretariat), carefully considered participant selection processes (e.g. host-driven, curated membership, blended model, or universal participation), collaborative agenda-setting mechanisms (e.g., multiple workstreams, coordinated succession, steering committee, or multi-year roadmaps), and a sustainable summit frequency (e.g., annual summits with ongoing engagement, flexible biannual summits, or annual summits with interim meetings). Any moves toward greater institutionalization must, however, preserve the flexibility that has been a hallmark of the series and avoid overly rigid structures that could hinder responsiveness to rapid technological change. Based on our analysis of key design elements, we favour:

**Table 8. Recommendations**

| Element | Recommended | Rationale |
| --- | --- | --- |
| 1. Hosting arrangement | Bidding system with regional rotation | Combines the benefits of competition (driving innovation and ensuring host commitment) with the goal of geographic diversity and inclusivity.<br><br>Ensures hosts have adequate resources and expertise.<br><br>Avoids potential pitfalls of a fixed rotation or purely regional approach. |
| 2. Secretariat format | Hybrid/incremental model | Strikes a balance between stability and flexibility.<br><br>Establishes a core team to provide continuity and institutional memory across summits, while allowing the host to retain significant influence.<br><br>Avoids the rigidity and potential bureaucracy of a fully independent secretariat, while addressing the shortcomings of a purely host-controlled model. |
| 3. Participant selection | Two-track approach | Ensures both focused, expert-driven discussions and wider engagement.<br><br>Track 1: Core group of AI-leading countries and major AI companies for in-depth discussions and decision-making on advanced AI governance.<br><br>Track 2: Broader participation for discussions on public interest opportunities and societal impacts of advanced AI. |
| 4. Agenda setting | Coordinated succession | Balances the need for continuity and long-term vision with host countries' interests in shaping the agenda.<br><br>Three consecutive hosts plan together ('troika') to ensure a coherent approach across summits.<br><br>Multi-year agenda provides a framework for discussions, while allowing flexibility to address emerging issues.<br><br>A steering committee provides guidance and expertise, while preserving host leadership. |
| 5. Meeting frequency | Annual summits with interim meetings | Balances high-level political engagement with in-depth technical discussions.<br><br>Annual summits provide a regular forum for decision-making and announcements.<br><br>Interim meetings allow for focused discussions on specific issues and scientific advancements among experts and ministers.<br><br>Ensures responsiveness to the rapid pace of AI development while allowing for sufficient preparation and follow-up between summits. |
| 6. Regular engagement with other fora | Regular engagement with UN and AISI Network. | Enhances the legitimacy and impact of both the summit series and the UN's efforts.<br><br>Gives the summit series a technical underpinning by engaging with the AISI Network.<br><br>Mechanisms could include aligning meeting schedules, sharing information, and ensuring complementarity between the summit series' focus on advanced AI, the UN's broader mandate on digital governance and the AISI Network's technical work. |

Implementation of these recommendations should proceed in phases:

**Table 9. Timeline for Implementation**

| Timeline | | Key Actions |
|---|---|---|
| **Short-term** (0 to 6 months) | | Designate hosts for the next three summits<br>Begin establishment of the secretariat structure |
| **Medium-term** (6-12 months) | | Formalize the participation framework<br>Implement the coordinated succession model<br>Create standardized handover procedures<br>Establish regular engagement with UN discussions |
| **Long-term** (12+ months) | | Evaluate and adjust the structure based on experience<br>Consider additional institutional support needs<br>Review and update long-term priorities |

The success of these changes will depend on maintaining the series' core strengths: its focused mandate, ability to drive action, and complementarity with other initiatives. Regular review and adjustment of these structures will ensure that they support rather than hinder the series' objectives.

Ultimately, the success of the AI Summit series will hinge on its ability to adapt and evolve while staying true to its core mission of fostering responsible development and deployment of advanced AI. The specific design choices outlined in this analysis offer options for strengthening the series' structure and operations. The summit series should continue to play a valuable role in shaping an international AI governance framework that maximizes the benefits of advanced AI while mitigating its risks, ultimately ensuring these powerful technologies serve the global public interest.

# 05. About the authors

**Lucia Velasco** is an economist and PhD researcher at the Oxford Martin AI Governance Initiative and the Maastricht School of Economics. She is a member of the Steering Committee and serves as Co-Chair of the Global Partnership on AI (GPAI-OECD). With over a decade of experience in the Spanish government, she has served as a policy advisor to several Ministers and Prime Ministers before becoming the first Director of the National Observatory for Technology and Society. Lucia has also worked with the World Economic Forum and the United Nations.

**Charles Martinet** is a Research Affiliate at the Oxford Martin AI Governance Initiative and Head of Policy at the French Center for AI Safety (CeSIA). Charles participates in the EU GPAI Code of Practice process as an independent expert, and was previously a Visiting Fellow at the Centre for the Governance of AI and a Talos Fellow. His work has been published by the OECD AI Policy Observatory, Euractiv, and the German Marshall Fund of the US, among others.

**Henry de Zoete** is a senior advisor and visiting fellow to the Oxford Martin AI Governance Initiative at the University of Oxford. Until July 2024 he was the Prime Minister's advisor on AI. He led the UK's approach on AI including the Bletchley Park AI Safety Summit and setting up the UK's AI Safety Institute. He was the UK's co-lead negotiator ("sherpa") for the AI Seoul Summit.

**Robert F. Trager** is Co-Director of the Oxford Martin AI Governance Initiative and Senior Research Fellow at the Blavatnik School of Government at the University of Oxford. He is a recognized expert in the international governance of emerging technologies, diplomatic practice, institutional design, and technology regulation. He regularly advises government and industry leaders on these topics.

**Duncan Snidal**, is a Professor of International Relations at the University of Oxford and professor emeritus at University of Chicago and the British Academy, researches problems of international cooperation and institutions–including international law and international organizations–with an emphasis on institutional design. His current projects focus on multi-partner governance of transnational production and the emergence of informal international organizations (such as the G20) as distinctive forms of international governance.

**Yoshua Bengio** is a Full Professor at Université de Montréal, and the Founder and Scientific Director of Mila – Quebec AI Institute. He earned the Association for Computing Machinery's 2018 A.M. Turing Award for contributions that shaped modern AI research.

**Ben Garfinkel** is a researcher at the University of Oxford and directs the Centre for the Governance of AI, a non-profit research and field-building organization focused on the opportunities and risks posed by AI. His research interests include the security implications of AI, the causes of war, and the methodological challenge of forecasting the impacts of emerging technologies.

**Kwan Yee Ng** is a Senior Program Manager at Concordia AI, a Beijing-based social enterprise focused on AI safety and governance. She writes for the International Scientific Report on the Safety of Advanced AI, a project overseen by 75 international AI experts and supported by 30 countries, the United Nations, and the European Union. Kwan Yee also serves as an affiliate at the Oxford Martin School's AI Governance Initiative.

**Haydn Belfield** is co-chair of the Global Politics of AI Project and a research fellow at the University of Cambridge's Leverhulme Centre for the Future of Intelligence. He has been a research associate and academic project manager at the University of Cambridge's Centre for the Study of Existential Risk for the past seven years.

**Don Wallace** works on policy development & strategy for Google DeepMind, focusing on the governance of frontier AI and AI for Science. He previously led DeepMind's UK policy work for over 5 years.

**Benjamin Prud'homme** is Vice-President of Policy, Safety and Global Affairs at Mila - Quebec AI Institute. He is an appointed expert of the OECD.AI Network, the United Nations Consultative Network of AI Experts, and UNESCO's AI Ethics Experts Without Borders.

**Brian Tse** is the Founder and CEO of Concordia AI, a social enterprise focused on AI safety and governance with presence in Beijing and Singapore.

**Roxana Radu** is an Associate Professor of Digital Technologies and Public Policy at the Blavatnik School of Government, University of Oxford.

**Ranjit Lall** is an Associate Professor of International Political Economy at the University of Oxford. His research focuses on the political economy of international cooperation, economic development, and technological change.

**Ben Harack** studies International Relations with a focus on the potential for AI to trigger major wars and ways that such outcomes can be avoided. His background includes experience with hardware and software as well as mathematics, physics, and psychology.

**Julia Morse** is a visiting fellow at the Oxford Martin AI Governance Initiative, a Robert A. Belfer International Affairs Fellow in European Security, and Assistant Professor of Political Science, University of California, Santa Barbara.

**Nicolas Miailhe** is co-founder & CEO of PRISM Eval, a Paris-based start-up specializing in red teaming, interpretability, and control of advanced AI models and systems. Founder & Chairman of the Board of the independent AI governance think-tank The Future Society (TFS); appointed expert on AI governance to the Global Partnership on AI (GPAI), OECD, and UNESCO.

**Scott Singer** is a Visiting Scholar at the Carnegie Endowment for International Peace, co-founder of the Oxford China Policy Lab, and affiliate of the Oxford Martin School AI Governance Initiative.

**Matt Sheehan** is a Fellow in the Asia Program at the Carnegie Endowment for International Peace. His research focuses on China's AI ecosystems and US-China AI issues.

**Max Stauffer** is the Co-CEO of the Simon Institute for Longterm Governance. Previously, Max served as Senior Science-Policy Officer at the Geneva Science-Policy Interface and as a Senior Visiting Fellow at UN University.

**Yi Zeng** is a Professor of AI at the Chinese Academy of Sciences. He is also the Founding Director of the Beijing Institute of AI Safety and Governance.

**Joslyn Barnhart** is a Senior Research Scientist at Google DeepMind working on long-term strategy and AI governance. Prior to Google DeepMind, she was Associate Professor of Government at Wesleyan University and faculty at University of California Santa Barbara focusing on international relations.

**Imane Bello (Ima)** is a lawyer specialized in AI compliance, international governance and risk management. She currently leads the Future of Life Institute's (FLI) effort on the AI Action Summit.

**Xue Lan** is Dean of the Institute for AI International Governance (I-AIIG) and Dean of Schwarzman College at Tsinghua University.

**Oliver Guest** is a research analyst at the Institute for AI Policy and Strategy. His research focuses on the international governance of advanced AI.

**Duncan Cass-Beggs** is executive director of the Global AI Risks Initiative at the Centre for International Governance Innovation (CIGI), focusing on developing innovative governance solutions to address current and future global issues relating to artificial intelligence. Duncan has over 25 years experience in public policy within government and international organizations.

**Lu Chuanying** is Deputy Director of the Institute for Cyberspace Governance Studies, China. His research focuses on Global AI governance and Cybersecurity.

**Sumaya Nur Adan** is an AI governance researcher with a background in Law and AI. She is currently a Research Affiliate at the Oxford Martin AI Governance Initiative, focusing on international organisations, AI risk assessment, institutional design, and benefit-sharing institutions.

**Markus Anderljung** is the Director of Policy and Research at the Centre for the Governance of AI (GovAI). He is also an Adjunct Fellow at the Center for a New American Security and a member of the OECD AI Policy Observatory's Expert Group on AI Futures.

**Claire Dennis** is a Research Scholar at the Centre for the Governance of AI, affiliate of the Oxford Martin School AI Governance Initiative, and Senior AI Adviser to the UN Center for Policy Research. Her background in diplomacy informs her work on global AI governance strategies across academic, policy, and international contexts.

**Acknowledgments**

The authors would also like to thank Sam Daws, Joe Jones, and José Villalobos for their valuable input and feedback.